
%
\input harvmac

\def\CTPa{\it Center for Theoretical Physics, Department of Physics,
      Texas A\&M University}
\def\CTPb{\it College Station, TX 77843-4242, USA}
\def\HARCa{\it Astroparticle Physics Group,
Houston Advanced Research Center (HARC)}
\def\HARCb{\it The Woodlands, TX 77381, USA}

\def\CERN{\it CERN Theory Division, 1211 Geneva 23, Switzerland}
\def\ie{\hbox{\it i.e.}}     
\def\eg{\hbox{\it e.g.}}

\catcode`\@=11 

\def\lsim{\mathrel{\mathpalette\@versim<}}
\def\gsim{\mathrel{\mathpalette\@versim>}}
\def\@versim#1#2{\vcenter{\offinterlineskip
    \ialign{$\m@th#1\hfil##\hfil$\crcr#2\crcr\sim\crcr } }}
\def\boxit#1{\vbox{\hrule\hbox{\vrule\kern3pt
      \vbox{\kern3pt#1\kern3pt}\kern3pt\vrule}\hrule}}

\def\etal{{\it et. al.}}
\def\r#1{$\bf#1$}
\def\rb#1{$\bf\overline{#1}$}

\def\t1{{\tilde 1}}

\def\JL{J. L. Lopez}
\def\DVN{D. V. Nanopoulos}

\def\eV{\,{\rm eV}}
\def\MeV{\,{\rm MeV}}
\def\GeV{\,{\rm GeV}}

\def\NPB#1#2#3{Nucl. Phys. B {\bf#1} (19#2) #3}
\def\PLB#1#2#3{Phys. Lett. B {\bf#1} (19#2) #3}

\def\PRD#1#2#3{Phys. Rev. D {\bf#1} (19#2) #3}
\def\PRL#1#2#3{Phys. Rev. Lett. {\bf#1} (19#2) #3}

\nref\GXa{GALLEX Collaboration, P. Anselmann \etal, preprint
GX 1-1992 (1992).}
\nref\HS{R. Davis, Jr., \etal, in Proceedings of the 21th International
Cosmic Ray Conference, Vol. 12, edited by R. J. Protheroe (University of
Adelaide Press, Adelaide, 1990), p. 143.}
\nref\KII{K. S. Hirata, \etal, \PRL{65}{90}{1297} and \PRL{66}{91}{9}.}
\nref\SAGE{A. I. Abazov, \etal, \PRL{67}{91}{3332}.}
\nref\BUTC{J. N. Bahcall and R. N. Ulrich, {\it Rev. Mod. Phys.} {\bf60} (1988)
297; S. Turck-Chieze, S. Cahen, M. Casse, and D. Doom, {\it Astrophysical J.}
{\bf335} (1988) 415.}
\nref\helio{D. B. Guenther, P. Demarque, Y.-C. Kim and M. H. Pinsonneault,
{\it Astrophys. J.} {\bf387} (1991) 377, but see D.R.O.Morrison, CERN preprint
PPE/92-109 (1992).}
\nref\BHKL{See, {\it e.g.,} S. Bludman, D. Kennedy, and P. Langacker,
\NPB{374}{92}{373} and \PRD{45}{92}{1810}.}
\nref\GXb{GALLEX Collaboration, P. Anselmann \etal, preprint GX 2-1992 (1992).}
\nref\MSW{L. Wolfenstein, \PRD{17}{78}{2369} and \PRD{20}{79}{2634};
S. P. Mikheyev and A. Yu. Smirnov, {\it Yad. Fiz.} {\bf42} (1985) 1441 and
{\it Nou. Cim.} {\bf9C} (1986) 17.}
\nref\Yanagida{T. Yanagida, {\it Prog. Theo. Phys.} {\bf B135} (1978) 66.}
\nref\Gellmann{M. Gell-Mann, P. Ramond, and R. Slansky, in {\it Supergravity},
ed. P. van Nieuwenhuizen and D. Freedman (North Holland, Amsterdam 1979),
p. 315.}
\nref\NIPS{M. Fukugita, M. Tanimoto, and T. Yanagida, Yukawa Institute
preprint YITP/K-983 (1992).}
\nref\many{R. Schaefer and Q. Shafi, Bartol preprint BA-92-28 (1992);
G. Efstathiou, J. R. Bond, and S. D. M. White, Oxford University preprint
OUAST/92/11 (1992); A. N. Taylor and M. Rowan-Robinson, Queen Mary College
preprint, (1992); M. Davis, F.J. Summers and D. Schlegel, Berkeley preprint
CfPA-TH-92-016 (1992).}
\nref\cobe{G. F. Smoot, \etal, COBE preprint (1992);
E. L. Wright, \etal, COBE preprint (1992).}
\nref\NO{\DVN\ and K. Olive, Nature {\bf327} (1987) 487.}
\nref\chorus{CHORUS collaboration, N. Armenise et al., CERN-SPSC/90-42 (1990).}
\nref\nomad{NOMAD collaboration, P. Astier et al., CERN-SPSC/91-21 (1991).}
\nref\fnal{FNAL proposal P803, K. Kodama et al., (1991).}
\nref\AEHN{I. Antoniadis, J. Ellis, J. Hagelin, and \DVN, \PLB{194}{87}{231}.}
\nref\GN{H. Georgi and \DVN, \PLB{82}{79}{392}, \NPB{155}{79}{52},
\NPB{159}{79}{16}.}
\nref\DHR{S. Dimopoulos, L. Hall, and S. Raby, \PRL{68}{92}{1984} and
\PRD{45}{92}{4192}.}
\nref\Barger{V. Barger, M. Berger, T. Han, and M. Zralek, \PRL{68}{92}{3394}.}
\nref\Kennedy{D. Kennedy, Fermilab preprint FERMILAB-PUB-92/149-T (1992).}
\nref\DL{L. Durand and \JL, \PRD{40}{89}{207}.}
\nref\Fritsch{H. Fritzsch, \PLB{70}{77}{436} and \PLB{73}{78}{317}.}
\nref\BKL{S. Bludman, N. Hata, D. Kennedy, and P. Langacker,
University of Pennsylvania preprint PUR-0516T (1992).}
\nref\EFL{J. Ellis, G. L. Fogli, and E. Lisi, \PLB{274}{92}{456}.}
\nref\KT{See \eg, E. Kolb and M. Turner, {\it The Early Universe}
(Addison-Wesley, 1990).}
\nref\NUO{R. Maschuw, Proc. Joint International Symposium and Europhysics
Conference on High Energy Physics, Geneva 1991, eds. S. Hegarty, K. Potter
and E. Quercigh (World Scientific, Singapore, 1992), vol. I, p. 619.}
\nref\Barr{S. Barr, \PLB{112}{82}{219}, \PRD{40}{89}{2457}; J. Derendinger,
J. Kim, and \DVN, \PLB{139}{84}{170}.}
\nref\neutrino{G. Leontaris, \PLB{207}{88}{447}; G. Leontaris and \DVN,
\PLB{212}{88}{327}; G. Leontaris and K. Tamvakis, \PLB{224}{89}{319};
S. Abel, \PLB{234}{90}{113}; I. Antoniadis, J. Rizos, and K. Tamvakis,
\PLB{279}{92}{281}.}
\nref\sharpening{\JL\ and \DVN, \PLB{268}{91}{359}.}
\nref\KLN{S. Kalara, \JL, and \DVN, \PLB{245}{90}{421}; \NPB{353}{91}{650}.}
\nref\decisive{J. L. Lopez and \DVN, \PLB{251}{90}{73}.}
\nref\CL{See, {\it e.g.,} M. Cvetic and P. Langacker, Univ. of
Pennsylvania preprint, UPR-505-T (1992).}

\Title{\vbox{\baselineskip12pt\hbox{CERN-TH.6569/92}\hbox{CTP--TAMU--53/92}
\hbox{ACT--15/92}}}
{\vbox{\centerline{The prospects for CHORUS and NOMAD}
\centerline{in the light of COBE and GALLEX}}}
\centerline{JOHN ELLIS$^{(a)}$, JORGE~L.~LOPEZ$^{(b)(c)}$, and
D.~V.~NANOPOULOS$^{(a)(b)(c)}$}
\smallskip
\centerline{$^{(a)}$\CERN}
\centerline{$^{(b)}$\CTPa}
\centerline{\CTPb}
\centerline{$^{(c)}$\HARCa}
\centerline{\HARCb}
\vskip .1in
\centerline{ABSTRACT}
The most natural MSW neutrino oscillation interpretation of the GALLEX and
other solar neutrino data, which invokes $m_{\nu_\mu}\sim3\times10^{-3}\eV$,
and
a general GUT see-saw hierarchy of neutrino masses, $m_{\nu_{e,\mu,\tau}}\sim
(m_{u,c,t})^2/M_U$, suggest that $m_{\nu_\tau}\sim10\eV$ in agreement with
the preference of COBE and other data on large-scale structure in the Universe
for a hot component in the Dark Matter. The general see-saw model also suggests
that neutrino mixing angles are related to quark mixing angles, which is also
consistent with the oscillation interpretation of the solar neutrino data, and
suggests that the forthcoming CHORUS and NOMAD experiments at CERN have a good
chance of observing $\nu_\mu - \nu_\tau$ oscillations. We present a minimal
realization of the general see-saw hierarchy in the context of flipped $SU(5)$.
\bigskip
\leftline{CERN-TH.6569/92}
\leftline{CTP-TAMU-53/92}
\leftline{ACT 15/92}
\leftline{June 1992}
\Date{}

\newsec{Introduction}
    The recent GALLEX data \GXa\ have added a fascinating new
twist to the continuing saga of solar neutrinos. They find a solar neutrino
deficit, as did the Homestake \HS\ and Kamioka \KII\ experiments, although
seemingly not as large a deficit as that reported by the SAGE experiment \SAGE.
The interpretation of these measurements is not yet clear, with explanations
being sought in nuclear physics -- are all the reaction rates correct? -- in
astrophysics -- is the standard solar model correct? -- and in particle physics
-- do neutrinos oscillate? We discard the nuclear hypothesis: it seems that the
residual rate uncertainties are no longer sufficient to make the deficit go
away \BUTC. More questionably, we also discard the astrophysical
hypothesis: simply reducing the core temperature of the Sun would suppress the
higher-energy Kamioka data more than the Homestake data, whereas the
opposite seems to be the case, and helioseismological observations
are by now severely constraining alternative solar models \refs{\helio,\BHKL}.

    We are left with the neutrino oscillation hypothesis, or rather
hypotheses, since there are several possible oscillation scenaria \GXb.
These include vacuum oscillations with $\Delta m_\nu^2\sim10^{-10}\eV^2$ and a
large mixing angle $\theta$, as well as two matter-enhanced
Mikheyev-Smirnov-Wolfenstein (MSW) \MSW\ possibilities with
$\Delta m_\nu^2\sim10^{-5}\eV^2$
and $\sin^22\theta > 1/2$ or $\sim10^{-2}$.
As discussed later, we find
large mixing angles theoretically implausible, and therefore focus here
on the $\Delta m_\nu^2\sim10^{-5}\eV^2$
and $\sin^22\theta\sim10^{-2}$ scenario.
Again as reviewed later, a general GUT see-saw hierarchy mechanism
\refs{\Yanagida,\Gellmann}
suggests that $m_{\nu_{e,\mu,\tau}}\sim(m_{u,c,t})^2/M_U$ so that
$m_{\nu_e}\ll m_{\nu_\mu} \ll m_{\nu_\tau}$, and also that
$\theta_{e\mu}\gg\theta_{e\tau}$, so we assume that the Sun is telling us about
$\nu_e - \nu_\mu$ oscillations.\foot{We discard the reports of
$\nu_e - \nu_\mu$ oscillations in the cosmic-ray neutrino flux, which the type
of model discussed here cannot reconcile with any of these solar neutrino
oscillation scenaria: see however \NIPS.} The see-saw mechanism therefore
suggests that $m_{\nu_\mu}\sim3\times10^{-3}\eV$ and hence that
$m_{\nu_\tau}\sim10\eV$.

    We are impressed by the concordance between this numerology and
the best-fit interpretation \many\ of COBE \cobe\ and other data on large-scale
structure in the Universe, which includes a hot dark matter (HDM)
component as well as the dominant cold dark matter (CDM) component.\foot{Such
a scenario was suggested by K. A. Olive and one of us (D.V.N.) many years
ago \NO.} Although this may involve taking the data too seriously, it does seem
as if the COBE data are most easily reconciled with the data on
peculiar velocities and the galaxy-galaxy correlation function at
large angles if $\Omega_{HDM}\sim0.3$ and $\Omega_{CDM}\sim0.7$ with $\Omega_B
0.1$ \many. If the $\nu_\tau$ constitutes this HDM, it should weigh several eV,
in agreement with the above estimate based on solar neutrino
oscillations and the see-saw mechanism.

    The purpose of this note is to link these remarks with the
prospects for observing $\nu_\mu - \nu_\tau$ oscillations in the CHORUS
\chorus\
and NOMAD \nomad\ experiments now being prepared at CERN. The above
``consensus" value of $m_{\nu_\tau}$ is certainly big enough for the CERN
experiments to detect $\nu_\mu - \nu_\tau$ oscillations if the $\nu_\mu -
\nu_\tau$ mixing angle $\theta_{\mu\tau}$ is large enough. In fact, the general
GUT see-saw mechanism links the neutrino mixing angles with the
corresponding neutrino masses, at least qualitatively. We show
that this theoretical expectation is consistent with the value of
$\theta_{e\mu}$ indicated by the GALLEX and other solar neutrino
experiments, and use it to estimate the magnitude of $\theta_{\mu\tau}$,
finding that it does indeed put $\nu_\mu - \nu_\tau$ oscillations within
reach of the CHORUS and NOMAD experiments, as well as the P803 proposal \fnal\
at FNAL. We illustrate these remarks with a minimal realization of the general
GUT see-saw mechanism in the framework of a generic field-theoretical flipped
SU(5) model \AEHN, showing explicitly how
the phenomenologically interesting mass ratios arise in this model.

\newsec{Review of the GUT see-saw mechanism}
Small neutrino masses are most naturally realized in terms of the GUT see-saw
mechanism \refs{\Gellmann,\Yanagida},
wherein the light left-handed neutrino
fields ($\nu$) of the
Standard Model interact with new superheavy right-handed (Majorana) neutrino
fields ($\nu^c$) through Dirac mass terms, as follows
\eqn\I{\bordermatrix{&\nu&\nu^c\cr \nu&0&m\cr \nu^c&m&M\cr}
\longrightarrow m_\nu\approx{m^2\over M},\quad m_{\nu^c}\approx M.}
Each of the entries $(m,M)$ in the above matrix should actually be regarded
as a $3\times3$ submatrix in generation space. The generic large mass-scale
$M$ is normally related to $M_U\approx10^{15}$ to $10^{19}\GeV$, and the
``0" entry may also be ${\cal O}(m^2/M)$ in some models.
The Dirac mass term above appears quite naturally in GUTs, when $\nu^c$ is
embedded in a suitable representation, such as a \r{1} of $SU(5)$ or a \r{16}
of $SO(10)$ \GN.
The difficulties and ambiguities arise when trying also to obtain
a Majorana mass term for $\nu^c$. This term does not arise in a minimal
$SU(5)$ GUT. In $SO(10)$ GUTs it can arise with the introduction of a \r{126}
representation \Gellmann\ with suitable vacuum expectation values and
additional
singlet fields \GN. The Dirac mass term in this case is generally given by the
up-quark mass matrix. It should be stressed though that the details of this
mechanism in $SO(10)$ GUTs tend to be rather complicated.
For the time being we will simply assume that the neutrino masses indeed
scale with the up-quark squared masses,\foot{In Sec. 4 we present a simple
realization of this statement based on the flipped $SU(5)$ gauge group \AEHN.}
that is, they obey the following mass hierarchy
\eqn\III{m_{\nu_e}:m_{\nu_\mu}:m_{\nu_\tau}\sim m^2_u:m^2_c:m^2_t,}
with some correction factors to be discussed later.
In principle, the see-saw mechanism would also predict the low-energy neutrino
mixings in analogy with the Cabibbo-Kobayashi-Maskawa (CKM) quark mixings.
However, this is a much more involved proposition, since the inter-generation
dependence of the $6\times6$ see-saw matrix needs to be known.

Here we simply assume that the heavy
neutrino degrees of freedom are integrated out and at low energies one ends up
with an effective $3\times3$ neutrino mass matrix $M_\nu$. Without loss of
generality, we can choose a basis where the charged lepton mass matrix is
diagonal and therefore the orthogonal matrix (assuming that $M_\nu$ is
symmetric) $V_\nu$, which diagonalizes $M_\nu$, contains the relevant mixing
angles. Since we expect these angles to be small, we can parametrize this
matrix as follows
\eqn\IVa{V_\nu\approx
\pmatrix{1&\theta_{e\mu}&0\cr\theta_{e\mu}&1&\theta_{\mu\tau}\cr
0&\theta_{\mu\tau}&1\cr},}
where we assume that the corresponding $\theta_{e\tau}$ angle is very small
(see below).

\newsec{Predictions for CHORUS and NOMAD}
The neutrino mass ratios in Eq. \III\ neglect the running of the
parameters between the unification scale $M_U$ and the low-energy quark mass
scale ($m_q$ or $1\GeV$, whichever is larger: below this scale the quark masses
do not run any more). The ratios in Eq. \III\ also assume a
generation-independent Majorana mass $M$, which is not necessarily true in
models, as we discuss later. Incorporating the running of the parameters and
relaxing the assumption of generation-independence one obtains \DHR
\eqn\IV{{m_{\nu_e}\over m_{\nu_\mu}}=\left[{\lambda_u(M_U)\over\lambda_c(M_U)}
\right]^2{M_2\over M_1}\approx\left[{m_u\over m_c}\right]^2{M_2\over M_1},}
where $m_u\equiv m_u(1\GeV)=5.1\pm1.5\MeV$ and $m_c\equiv m_c(m_c)=1.27\pm0.05
\GeV$, use has been made of the common running of the up- and charm-quark
Yukawa couplings, and we assume $M\equiv{\rm diag}(M_1,M_2,M_3)$. Also,
\eqn\V{{m_{\nu_\mu}\over
m_{\nu_\tau}}=\left[{\lambda_c(M_U)\over\lambda_t(M_U)}
\right]^2{M_3\over M_2}=\left[{\lambda_c(m_t)\over\lambda_t(m_t)}\right]^2
\left[1-{\lambda^2_t(m_t)\over\lambda^2_C}\right]{M_3\over M_2}=
\left[{m_c\over\eta_c m_t}\right]^2\left[1-\left({m_t\over190}\right)^2\right]
{M_3\over M_2},}
where $\eta_c\equiv m_c(m_c)/m_c(m_t)\approx1.9$ for a plausible range of
values of $\alpha_3(M_Z)$ \refs{\DHR,\Barger,\Kennedy}, and $\lambda_C$ is the
critical value of the top-quark Yukawa coupling above which $\lambda_t$
develops a Landau pole below $M_U$; one obtains
$\lambda_t(m_t)/\lambda_C\approx m_t/190\GeV$. The value of $\lambda_C$ is
decreased by $\lsim5\%$ if the bottom-quark Yukawa coupling is not neglected.
In Eq. \V\ we have used the following approximate formulae \refs{\Kennedy,\DL}
\eqna\Va
$$\eqalignno{\lambda^2_c(m_t)&=\eta\lambda^2_c(M_U),&\Va a\cr
\lambda^2_t(m_t)&=\eta\left[1-{\lambda^2_t(m_t)\over\lambda^2_C}\right]
\lambda^2_t(M_U),&\Va b\cr}$$
where $\eta\equiv\prod_{i=1}^3(\alpha_U/\alpha_i)^{c_i/b_i}$ with
$c_i=({13\over15},3,{16\over3})$ and $b_i=({33\over5},1,-3)$. Note the
significant modification of the $m_{\nu_\mu}/m_{\nu_\tau}$ ratio due to the
running of parameters involved.

For the neutrino mixing angles we make the following
phenomenologically-motivated ans\"atze
\eqn\VI{\theta_{e\mu}=(m_{\nu_e}/m_{\nu_\mu})^{1/4}\quad{\rm and}\quad
\theta_{\mu\tau}=(m_{\nu_\mu}/m_{\nu_\tau})^{1/2},}
which resemble predictions for the CKM angles based on certain textures for the
quark mass matrices \Fritsch ,\GN.
Indeed, from Eqs. \IV\ and \V\ one can verify
that $\theta_{\mu\tau}\sim\theta^2_{e\mu}$ for a plausible range of the
parameters, whilst $\theta_{e\tau}$ would be much smaller.
Therefore, $V_\nu$ in Eq. \IVa\ has the
same texture as the CKM matrix, although with $\theta_{e\mu}\ll\theta_c$.
Accounting for all the proper factors, the relations \VI\ give
\eqn\VII{\sin^22\theta_{e\mu}=4\left[m_u\over m_c\right]
                                \left(M_2\over M_1\right)^{1/2},}
and
\eqn\VIII{\sin^22\theta_{\mu\tau}=4\left[{m_c\over\eta_c m_t}\right]^2
\left[1-\left({m_t\over190}\right)^2\right]{M_3\over M_2},}
which we now confront with experiment.

The prediction for $\sin^22\theta_{e\mu}$ in Eq. \VII\ gives a central
value of $1.6\times10^{-2}(M_2/M_1)^{1/2}$. Since the best fits to the
GALLEX data in terms of the MSW mechanism give
$\Delta m^2_{\nu_e-\nu_\mu}=(0.3-1.2)\times 10^{-5}\eV^2$ and
$\sin^22\theta_{e\mu}=(0.4-1.5)\times10^{-2}$ \refs{\GXb,\BKL}, we deduce
that $M_2/M_1\lsim1$ is required.

The predictions for the $\mu-\tau$ sector follow from Eqs. \V\ and \VIII.
These are given in Table I for central values of the parameters and
$m_{\nu_\mu}=2\times10^{-3}\eV$ (where relevant). The ratio $M_3/M_2$ is
an additional free parameter, although theoretically we expect (see Sec. 4)
$1\lsim M_3/M_2\lsim100$: we took $M_3/M_2=10$ in Table I.
The last column gives the cosmic relic density of $\tau$-neutrinos, where $h$
is the Hubble parameter ($0.5\le h \le1$). This quantity is simply given by
$\Omega_\nu h^2=m_\nu/91.5\eV$ \KT. We find it truly amazing that several
apparently unrelated experimental measurements and theoretical estimates
-- solar neutrinos, the MSW mechanism, the see-saw mechanism, and estimates
of $m_t$ converge to give $\Omega_\nu\sim0.3$ in agreement with the COBE data
for $h\sim0.5$ to 1. Either this is a remarkable and vicious coincidence, or
$\ldots$.

\topinsert
\noindent {\bf Table I}: Values of the $\nu_\mu-\nu_\tau$ mixing parameters,
$\tau$-neutrino mass, and relic cosmological density as a function of the
top-quark mass \EFL\ for the model discussed in Sec. 3 and central values of
the
parameters (including $M_3/M_2=10$). Where relevant we have assumed that
$m_{\nu_\mu}=2\times10^{-3}\eV$.
\medskip
\input tables
\thicksize=1.0pt
\centerjust
\begintable
$m_t\,(\!\GeV)$|$\sin^22\theta_{\mu\tau}$|
$\Delta m^2_{\nu_\mu-\nu_\tau}\,(\!\eV^2)$|$m_{\nu_\tau}\,(\!\eV)$|
$\Omega_\nu h^2$\cr
$90$|$1.7\times10^{-3}$|$22$|$4.7$|$0.05$\nr
$100$|$1.3\times10^{-3}$|$38$|$6.2$|$0.07$\nr
$110$|$9.8\times10^{-4}$|$66$|$8.1$|$0.09$\nr
$120$|$7.5\times10^{-4}$|$115$|$11$|$0.12$\nr
$130$|$5.6\times10^{-4}$|$202$|$14$|$0.16$\nr
$140$|$4.2\times10^{-4}$|$368$|$19$|$0.21$\nr
$150$|$3.0\times10^{-4}$|$715$|$27$|$0.29$\endtable
\bigskip
\endinsert

The current limits \NUO\
on $\nu_\mu-\nu_\tau$ oscillations exclude values of
$\sin^22\theta_{\mu\tau}>4\times10^{-3}$ for $\Delta m^2_{\nu_\mu-\nu_\tau}
\gsim50\eV^2$, with considerably weaker upper bounds for smaller values of
$\Delta m^2_{\nu_\mu-\nu_\tau}$. The proposed CHORUS \chorus\ and NOMAD \nomad\
experiments at CERN and P803 \fnal\ at Fermilab plan to probe values of
$\sin^22\theta_{\mu\tau}$
one order of magnitude lower than the current experimental upper bound, with
a similar sensitivity to $\Delta m^2_{\nu_\mu-\nu_\tau}$. This means (see
Table I) that the predictions for the simple model presented in this section
should be fully testable by these new experiments.

Note that since $m_{\nu_\tau}\propto(M_3/M_2)^{-1}$, $M_3/M_2\sim1$ would give
a $\tau$-neutrino relic density a factor of 10 larger than the values given in
Table I and therefore in conflict with current cosmological observations, which
require $\Omega_\nu<1$ and appear to favor $\Omega_\nu\sim0.3$. Conversely,
values of $M_3/M_2\sim100$ would make the $\tau$-neutrinos cosmologically
uninteresting, even though their mixing with $\mu$-neutrinos would be enhanced
by a factor of 10 relative to Table I, but still unconstrained
experimentally \NUO\
due to the smallness of $\Delta m^2_{\nu_\mu-\nu_\tau}$ (a
factor of 100 smaller than in Table I).

\newsec{The flipped see-saw mechanism}
We now describe an underlying see-saw mechanism which can produce the above
phenomenologically interesting neutrino mass ratios. We do this in the simplest
unified supersymmetric extension of the Standard Model which predicts non-zero
neutrino masses, namely in the context of flipped $SU(5)$
\refs{\Barr,\AEHN}.
In this model the see-saw matrix \refs{\neutrino,\sharpening}
for each generation involves three fields:
$\nu_i,\nu^c_i,\phi_i$, where $\phi_i$ is an $SU(5)\times U(1)$ singlet field,
as follows
\eqn\IX{\bordermatrix{&\nu_i&\nu^c_i&\phi_i\cr
\nu_i&0&m_{ui}&0\cr \nu^c_i&m_{ui}&\lambda_{9i}M^2_U/M_{nr}&\lambda_{6i}M_U\cr
\phi_i&0&\lambda_{6i}M_U&\mu_i\cr},}
where the various entries come from the following flipped $SU(5)$ couplings
\eqna\X
$$\eqalignno{&\lambda_{ui}F_i\bar f_i\bar h\to m_{ui}\nu_i\nu^c_i,&\X a\cr
&\lambda_{6i}F_i\bar H\phi_i\to \lambda_{6i}\bar V\nu^c_i\phi_i
\approx\lambda_{6i} M_U\nu^c_i\phi_i,&\X b\cr
&\lambda_{9i}{1\over M_{nr}}F_iF_i\bar H\bar H\to
\lambda_{9i}{\bar V^2\over M_{nr}}\nu^c_i\nu^c_i
\approx\lambda_{9i}{M^2_U\over M_{nr}}\nu^c_i\nu^c_i,&\X c}$$
and the $\mu_i\phi_i\phi_i$ mass term. In these expressions, $F_i,\bar f_i$
are the usual \r{10},\rb{5} matter fields, $H,\bar H$ are the \r{10},\rb{10}
$SU(5)\times U(1)$ breaking Higgs representations whose neutral components
$(\nu^c_H,\nu^c_{\bar H})$ acquire vacuum expectation values $(V=\bar V\approx
M_U)$, and $M_{nr}\approx10^{18}\GeV$ is
the scale of calculable \KLN\
non-renormalizable terms
in the superpotential \decisive. The calculable
non-renormalizable couplings are a
feature of string-derived flipped $SU(5)$ models
that we expect on general
grounds \refs{\KLN,\decisive}
to have counterparts in other string-derived models \CL.
The light eigenvalue of the $3\times3$ see-saw matrix is simply given by
\eqn\XI{m_{\nu_i}\approx{m^2_{ui}\over M_i},}
with
\eqn\XII{M_i=\lambda^2_{6i}{M^2_U\over \mu_i}\left(1-{\lambda_{9i}\over
\lambda^2_{6i}}{\mu_i\over M_{nr}}\right).}
Neglecting the higher-order term (\ie, setting $\lambda_{9i}\equiv0$) we find
\eqn\XIII{{M_3\over M_2}={\lambda^2_{63}\over\lambda^2_{62}}{\mu_2\over\mu_3}.}
In string models it is quite common to have a hierarchical set of Yukawa
couplings, in that usually only the third generation gets ${\cal O}(1)$
couplings; the first and second generation Yukawa couplings are suppressed
by powers of $\mu/M_{nr}\sim1/10$ \decisive. The masses $\mu_i$ may or may not
obey any such hierarchy. In specific models we then expect $\lambda_{62}
/\lambda_{63}\sim1/10$ and therefore $M_3/M_2\sim1-100$ if we also allow for
a possible hierarchy in the $\mu_i$. We note in passing the key role of the
$\lambda_6$ couplings \X{b}, which are allowed (even compulsory) in flipped
$SU(5)$, but whose phenomenological importance has been hidden until now.

Let us now see if a $\tau$-neutrino mass $\sim10\eV$ can be obtained in this
model. From Eq. \XI\ we have
\eqn\XIV{m_{\nu_\tau}=
{m^2_t(M_U)\over M_3}={m^2_t\over{\eta\left[1-(m_t/190)^2\right]}}{1\over
M_3},}
which gives $m_{\nu_\tau}\sim1-10\eV$ for $M_3\sim10^{12}\GeV$ (with
$\eta\approx10$). Also, from Eq. \XII\ $M_3\sim\lambda^2_{63}M^2_U/\mu_3
\sim10^{12}\GeV$ for $M_U\sim10^{15}\GeV$, $\mu_3\sim10^{17}\GeV$, and
$\lambda_{63}\sim1/3$, all perfectly reasonable numbers.

\newsec{Summary}
We have emphasized in this note that the MSW \MSW\ interpretation \GXb\ of
the GALLEX solar neutrino data \GXa, extrapolated by a general GUT see-saw
mechanism \refs{\Yanagida,\Gellmann},
is consistent with the suggestion from
COBE \cobe\ that there may be a hot component in the Dark Matter \many, namely
a $\nu_\tau$ weighing ${\cal O}(10\eV)$. Furthermore, a plausible texture of
neutrino mixing angles, also motivated by GALLEX, suggests that
$\nu_\mu-\nu_\tau$ mixing would be observable in the CHORUS \chorus\ and
NOMAD \nomad\ neutrino oscillation experiments now being prepared at CERN.
We have also presented a specific realization of this general see-saw
mechanism in the context of flipped $SU(5)$ \refs{\Barr,\AEHN}.

We realize that neither the MSW interpretation of the GALLEX data nor the
cocktail interpretation of the COBE data are at all sure, let alone a GUT
see-saw mechanism. Nevertheless, we find these convergent indications
impressive, and hope they encourage our experimental colleagues.

\bigskip
\bigskip
\bigskip
\noindent{\bf Acknowledgements}: This work has been supported in part by DOE
grant DE-FG05-91-ER-40633. The work of J.L. has been supported in part by an
ICSC-World Laboratory Scholarship. The work of D.V.N. has been supported in
part by a grant from Conoco Inc.

\listrefs
\bye